\documentclass[journal]{journal}
\ifCLASSINFOpdf
 \usepackage[pdftex]{graphicx}
\else
\fi
\usepackage[cmex10]{amsmath}
\usepackage{amsfonts}

\usepackage{algorithmic}
\usepackage[ruled,vlined]{algorithm2e}

\SetCommentSty{mycommfont}

\SetKwInput{KwInput}{Input}                % Set the Input
\SetKwInput{KwOutput}{Output}              % set the Output
\usepackage[tight,footnotesize]{subfigure}
\usepackage{float,epsfig}

\usepackage{url}
\DeclareRobustCommand*{\IEEEauthorrefmark}[1]{%
  \raisebox{0pt}[0pt][0pt]{\textsuperscript{\footnotesize #1}}%
}
\begin{document}
\title{Deep Reinforcement Learning Approach for Trading Automation in The Stock Market}

\author{\IEEEauthorblockN{Taylan Kabbani\IEEEauthorrefmark{1},
Ekrem Duman \IEEEauthorrefmark{2}}\\
\IEEEauthorblockA{\IEEEauthorrefmark{}Department of Industrial Engineering, Ozyegin University, Istanbul, Turkey}\\
\IEEEauthorblockA{\IEEEauthorrefmark{1}taylan.kabbani1@ozu.edu.tr}\\
\IEEEauthorblockA{\IEEEauthorrefmark{2}ekrem.duman@ozyegin.edu.tr}\\
}


\markboth{Journal of \LaTeX\ Class Files,~Vol.~6, No.~1, January~2007}%
{Shell \MakeLowercase{\textit{et al.}}: Bare Demo of IEEEtran.cls for Journals}



\maketitle
\thispagestyle{empty}

\begin{abstract}
%\boldmath
Deep Reinforcement Learning (DRL) algorithms can scale to previously intractable problems. The automation of profit generation in the stock market is possible using DRL, by combining the financial assets price "prediction" step and the "allocation" step of the portfolio in one unified process to produce fully autonomous systems capable of interacting with its environment to make optimal decisions through trial and error. This work represents a DRL model to generate profitable trades in the stock market, effectively overcoming the limitations of supervised learning approaches. We formulate the trading problem as a Partially Observed Markov Decision Process (POMDP) model, considering the constraints imposed by the stock market, such as liquidity and transaction costs. We then solve the formulated POMDP problem using the Twin Delayed Deep Deterministic Policy Gradient (TD3) algorithm reporting a 2.68 Sharpe Ratio on unseen data set (test data) . From the point of view of stock market forecasting and the intelligent decision-making mechanism, this paper demonstrates the superiority of DRL in financial markets over other types of machine learning and proves its credibility and advantages of strategic decision-making.
\end{abstract}


% \fi
\IEEEpeerreviewmaketitle

\section{INTRODUCTION}

\IEEEPARstart{T}{he} prime objective of any investor when investing in any financial market is to minimize the risk involved in the trading process and maximize the profits generated. Investors can meet this objective by successfully predicting the prices or trends of the market assets and optimally allocating the capital among the selected assets. This process is very challenging for a human to consider all relevant factors in a complex and dynamic environment; therefore, the design of adaptive automated trading systems capable of meeting the investor's objective and bringing more stagnant wealth into the global market has been an intensive research topic. Many efforts have been made to design such trading systems in the past decade. The majority of these efforts focused on using \textit{Supervised learning} (SL) techniques \cite{PATEL2015259, 8010701, 10.1371/journal.pone.0234107, app10113961, 8489208}, which in essence train a predictive model (e.g., Neural Network, Random Forest,...) on historical data to forecast the trend direction of the market. Regardless of their popularity, these techniques suffered from various limitations, leading to sub-optimal results \cite{LopezdePrado2018The1R}.
\textit{Reinforcement Learning} (RL) offers to solve the drawbacks of Supervised Learning approaches in trading financial markets by combining the financial assets price "prediction" step and the "allocation" step of the portfolio in one unified process to optimize the objective of the investor, where the trading agent (the algorithm) interacts with the environment (the model) to take the optimal decision \cite{Meng}. In addition, financial data is highly time-dependent (function of time), making it a perfect fit for Markov Decision Processes (MDP) \cite{MDP_book}, which is the core process of solving RL problems. MDP captures the entire past data and defines the whole history of the problem in just the agent's current state, and that's highly crucial when it comes to modeling financial market data \cite{chakraborty2019capturing}.

Most works that studied the RL's applications in financial markets and particularly in trading stocks considered discrete action spaces \cite{8489208, Bertoluzzo2014, TAN20114741, Deng}, i.e., buy, hold, and sell a fixed number of shares to trade a single asset. In this work, a continuous action space approach is adopted to give the trading agent the ability to gradually adjust the portfolio's positions with each time step (dynamically re-allocate investments), resulting in better agent-environment interaction and faster convergence of the learning process. In addition, the approach supports the managing of a portfolio with several assets instead of a single one.
We first propose a novel formulation of the stock trading problem or what is referred to as the trading \textit{Environment} as a Partially Observed Markov Decision Process (POMDP) model considering the constraints imposed by the stock market, such as liquidity and transaction costs. More specifically, we design an environment that simulates the real-world trading process by augmenting the state (observation) representation with ten different technical indicators and sentiment analysis scores of news releases along with other state components. We then solve the formulated POMDP problem using the Twin Delayed Deep Deterministic Policy Gradient (TD3) algorithm, which can learn policies in high-dimensional and continuous action spaces like those typically found in the stock market environment. Finally, we evaluate our proposed approach by performing back-testing, which is the process used by traders and analysts to assert the viability of a trading strategy by testing it on historical data.

\section{BACKGROUND AND RELATED WORK}
\subsection{MDP in Reinforcement Learning}
In essence, \textit{Markov Decision Processes} \cite{alagoz_hsu_schaefer_roberts_2009} (MDP) is used to model stochastic processes containing random variables, transitioning from one state to another depending on certain assumptions and definite probabilistic rules. 
MDPs are a perfect mathematical framework to describe the reinforcement learning problem. In this framework, researchers call the learner or decision maker the \textit{agent} and the surrounding which the agent interacts with (comprising everything outside the agent) the \textit{environment}. The learning process ensues from the agent-environment interaction in MDP, at each time step $t \in \{1, 2, 3,..., T\}$ the agent receives some representation (information) of its current state from the environment $s_{t} \in \mathcal{S}$, and on that basis selects an action $a_{t} \in \mathcal{A}$ to perform. One step later, due to its action, the agent finds itself in a new state, and the environment returns a reward $R_{t+1} \in \mathcal{R}$ to the agent as a feedback of its action's quality \cite{RL_book}.

\subsection{The Objective of Reinforcement Learning}
The \textit{objective} of any RL problem is to maximize the cumulative reward $\mathbb{G}_{t}$ it receives in the long run instead of the immediate reward $R_{t}$
\begin{equation}
\label{Eq.reward}
	\mathsf{E}[\mathbb{G}_{t}] =\mathsf{E}[R_{t+1} + R_{t+2} + R_{t+3} + ... + R_{T}] 
\end{equation}
In the above reward equation (Eq. \ref{Eq.reward}), the term $R_{T}$ denotes the reward received at the terminal state T. meaning that the aforementioned equation  is only valid when the problem at hand is an \textit{Episodic task}, i.e., ends in a terminal state T. For the \textit{Continuous tasks} i.e., no terminal state, $T = \infty$, a discount factor gamma is introduced to Eq. \ref{Eq.reward} ($0 \leq \gamma \leq 1$):
$$\mathbb{G}_{t} = R_{t+1} +  \gamma R_{t+2} + \gamma^{2} R_{t+3} + ... + \gamma^{k-1} R_{t+k} + ....$$
\begin{equation}
\label{Eq.discount}
= \sum_{0}^{\infty} \gamma^{k} R_{t+k+1}
\end{equation}

\subsection{Bellman Equations}
\textit{Value functions} are being used by almost all RL methods to estimate how good (in terms of expected return) it is for the agent to be in a given state or to perform an action in a given state. This evaluation is being made based on the future expected sum of rewards. Accordingly, value functions are determined with respect to the future actions the agent will take. We call a particular way of acting a \textit{Policy} ($\pi$) \cite{RL_book} which is a function that maps from environment’s states to probabilities of selecting each possible action.

\textit{Bellman equations} \cite{bellman} are the fundamental property of value functions used in \textit{dynamic programming} as well as in reinforcement learning to solve MDPs, and they are essential to understand how many RL algorithms work. Bellman equation states that the value function of state s ($\mathcal{V}_{\pi}(s)$) can be calculated by finding the sum over all possibilities of expected returns, weighting each by its probability of occurring following a policy $\pi$:
\begin{equation}
\label{state_value}
   \mathcal{V}_{\pi}(s) \doteq \sum_{a}\pi(a|s)  \sum_{s^{\prime}} \sum_{r}
    							 P(s^{\prime},r|s,a)[r + \gamma \mathcal{V}_{\pi}(s^{\prime})], \forall s \in \mathcal{S}
\end{equation}
In a similar way we define the action-value ($q_{\pi}(s,a)$) function as:
\begin{equation}
\label{action_value}
	q_{\pi}(s,a) =  \sum_{s^{\prime}} \sum_{r}P(s^{\prime},r|s,a)[r + \gamma \sum_{a^{\prime}}\pi(a^{\prime}|s^{\prime}) q_{\pi}(s^{\prime},a^{\prime})]
\end{equation}
From Bellman equations (Eq. \ref{state_value} and Eq. \ref{action_value}) we can derive what is called \textit{The Bellman Optimality Equations}. Intuitively, the Bellman optimality equation expresses the fact that the value of a state under an optimal policy ($\pi_{*}$) must equal the expected return for the best action from that state \cite{RL_book}, and the optimal state-value function ($\mathcal{V}_{*}$) equals to :
\begin{equation}
\mathcal{V}_{*}(s) = \max_{a} \sum_{s^{\prime}}\sum_{r}P(s^{\prime},r | s, a)[r + \gamma \mathcal{V}_{*}(s^{\prime})]
\end{equation}
Similarly, we define optimal action-value ($q_{*}$) function as:
\begin{equation}
  \begin{aligned}
q_{*}(s,a) &= \max_{\pi} q_{\pi}(s,a) = \sum_{s^{\prime}}\sum_{r}P(s^{\prime}, r|s,a)[r+\gamma \max_{a^\prime}q_{*}(s^{\prime},a^{\prime})]
  \end{aligned}
\end{equation}

\subsection{Taxonomy of RL Algorithms}
RL algorithms are classified based on how to represent and train the agent into three main approaches:

\subsubsection{Critic-Only Approach}
\begin{figure}[htp]
\begin{center}
\includegraphics[width=\linewidth]{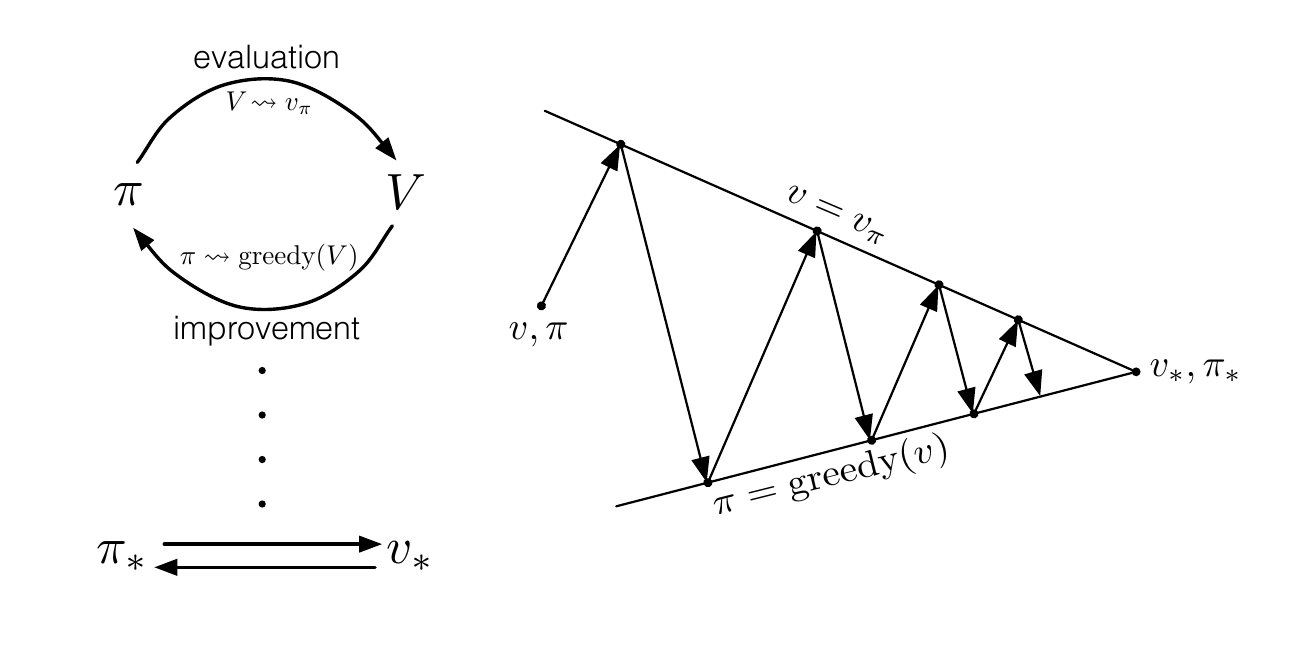}
\end{center}
\vspace*{-0.2cm}
\caption{Generalized Policy Iteration \cite{RL_book}}
\label{GPI_fig}
\end{figure}

This family of algorithms learns to estimate the value function (State-value function or action-value function) by using what is called \textit{Generalized Policy Iteration} (GPI), this concept refers to the interaction of two steps. The first step is the \textit{policy-evaluation}, the main goal of this step is to collect information (value functions) under the given policy to determine how good it is. The second step is the \textit{policy-improvement}, it is responsible of improving the policy by choosing greedy actions with respect to the value functions computed from the policy-evaluation step. The two steps alternate in a consecutive manner until the value functions and policies stabilize which means that the process has reached an optimal policy as illustrated in Fig. \ref{GPI_fig}.

We distinguish between two different ways the agent learns the value function of the system. The first way is \textit{Tabular Solution Method} where the value functions are represented as arrays or tables and updated with more accurate values after each iteration as the agent collects more experience. This way of learning often finds exact solutions. However, it does not generalize well, and the state and action spaces must be small enough to be stored in tables.

The second possible way in the critic-only approach is called \textit{Approximate Solution Method}, which tends to generalize better than Tabular Method but has lower discrimination, and it is capable of learning the value function of systems with enormous state and action spaces. Approximate methods achieve this generalization by combining RL with \textit{supervised learning} algorithms. Deep Reinforcement Learning is considered an approximate method that combines Neural Networks with RL. \textit{Mnih et al. (2013)} \cite{DQL_paper} is considered the father of DRL, where he trained an agent of Deep Q-network (DQN) to play Atari games, where pixels of the game screen were the input data (state), and the directions of the joystick were actions. He proved that DRL had outperformed all existing algorithms in 2015 \cite{Mnih_2015}.

\subsubsection{Actor-Only Approach}
All methods under the Critic-Only approach rely on the GPI framework to learn approximate action values to infer a good policy.  \textit{Actor-Only methods} (also called \textit{Policy Gradient Methods}) estimate the gradient of the objective which is maximizing rewards with respect to the policy parameters and adjust the policy parameters $\theta$ based on the estimate (Eq. \ref{theta}). The parameterized policy function takes state and action as an input and returns the probability of taking that action in that state instead of taking the state only as an input and returning the value function as Critic-Only methods do. Note that in the bellow equation $G_t$ represents the expected reward at time t.

\begin{equation}
\label{theta}
\theta_{t+1} = \theta_t + \alpha \nabla \ln\pi(a_t|s_t, \theta_t) G_t
\end{equation}

\subsubsection{Actor-Critic Approach}
In the \textit{Actor-Critic approach}, the actor's job is to select actions at each time step to form the policy, where the critic's role is to evaluate these actions taken by the actor. So the approach is gradually adjusting the policy parameters $\theta$ of the actor to take actions that maximize the total reward predicted by the critic.
The TD error ($\delta$) calculated by the critic to evaluate the action is as follows:
\begin{equation}
\label{delta}
\delta=  R_{t+1} +\gamma \hat{V}(s_{t+1}, w) - \hat{V}(s_{t}, w)
\end{equation}
The value function estimation of the current state $\hat{V}(s_{t}, w)$ is added as a baseline to make the learning faster.
The parameter $\theta$ of the actor is being adjusted in the way of maximizing the total future reward from Eq. \ref{theta} and Eq. \ref{delta} we conclude the equation used by the Actor-Critic to update the gradient at each time step t as the following:
\begin{equation*}
\theta_{t+1} = \theta_t + \alpha \nabla \ln\pi(a_t|s_t, \theta_t) (R_{t+1} +\gamma \hat{V}(s_{t+1}, w) - \hat{V}(s_{t}, w))
\end{equation*}

	Many researchers worked on improving the DQN algorithm. \textit{Van Hasselt et al. (2015)} \cite{Double_Q} proposed to use two networks instead of one Q-network to choose the action and the other to evaluate the action taken to solve the deviation problem in DQN. They called it Double-DQN. \textit{Lillicrap et al. (2018)} \cite{lillicrap2019continuou} built on the top of Double-DQN, an algorithm based on the deterministic policy gradient (DDPG) that can operate over continuous action spaces. The Twin Delayed Deep Deterministic Policy Gradient (TD3) algorithm which will be discussed in section \ref{TD3_sect}, was proposed by \textit{Fujimoto et al. (2019)} \cite{TD3} to tackle the problem of the approximation error in DDPG.
	
\subsection{RL in Finance}
\textit{Bertoluzzo and Corazza (2012)} \cite{Bertoluzzo2012} investigated the performance of different RL algorithms in day-trading for one selected Italian stock. Specifically, they compared the performance of Q-learning, and Kernal-based reinforcement learning, concluding that Q-learning performance outperformed Kernal-based RL. In a subsequent study (2014) \cite{Bertoluzzo2014}, they explored the effect of different reward functions such as Sharpe ratio, average log return, and OVER ratio on the performance of Q-learning. By trading six selected Italian stocks, they reported that lagged return reward function has the best performance. Instead of approximating a value function (critic-only), \textit{Deng et al. (2017)} \cite{Deng} made one of the first attempts on combining Deep Learning with Recurrent Reinforcement Learning to directly approximate a policy function. This approach is called ``deep recurrent reinforcement learning'' (DRRL). In their proposed method, first, the DL part extracts 45 useful features from the market to be used as state representative in the environment. Secondly, they use a Recurrent Neural Network (RNN) as a trading agent to interact with the deep-generated state features and make decisions. To investigate the potential advantage of Actor-Critic methods in solving the day trading problem, \textit{Conegundes and Pereira} (2020) \cite{beating_the_stock_market} used Deep Deterministic Policy Gradient (DDPG) algorithm to solve the asset allocation problem. Considering different constraints such as liquidity, latency, slippage, and transaction costs, they back-tested their approach on the Brazilian Stock Exchange datasets. They showed that their approach successfully obtained 311\% cumulative return in three years with an annual average maximum drawdown around 19\%.

\section{PROBLEM DESCRIPTION}
The stock trading problem is being modeled as \textit{Partially Observed Markov Decision Process} (POMDP), which can be formulated by describing its State Space, Action Space, and Reward Function. The POMDP model of the problem is called the trading environment, and it's built to carefully mimic the real-world trading process.
\subsection{State Space}
\label{StateSpace_sect}
The state-space in the proposed environment is designed to support multiple and single stock trading by representing the state as (1+ 13 x $\mathcal{N}$)-dimensional vector where $\mathcal{N}$ is the number of assets we consider to trade in the market. Hence the state space increases linearly with the number of assets available to be traded.

There are two main parts of the state presentation. The first part is the \textit{Position State} $\in \mathbb{R}^{1+\mathcal{N}}_{+}$  which holds the current cash balance and shares owned of each asset in the portfolio, and the second part of the state is the \textit{Market Signals} $\in \mathbb{R}^{12 \times \mathcal{N}}$, which holds the necessary market features for each asset as a tuple, these features are the required information provided to the agent to make predictions of the market movement. The first type of information is based on the hypothesis of \textit{technical analysis} \cite{technical_analysis}, which states that the future behavior of financial markets is conditioned on its past; hence technical indicators are being used in the state space to help the agent interpret the market behavior. The second type of information is based on \textit{fundamental analysis} \cite{fundamental_analysis}, which studies everything from the overall economy and industry conditions to news releases. Therefore a Natural Language Processing (NLP) approach is used to measure the general sentiment from the news releases and integrate it with the state representation. The state (observation) vector at each time step is provided to the agent as follows:

$$\mathbf{S_t} = \mathbf{[[b_t, h_t], [\{(C^i_t,  SS^i_t, T^i_t) | i \in \mathcal{N}\}]]}$$

Each component of the state space is defined as follows:
\begin{itemize}
\item $\mathcal{N} \in \mathbb{Z}^{\mathcal{N}}_+$: Number of assets in the portfolio.
\item $\mathbf{b_t} \in \mathbb{R}_+$: The available cash balance in the portfolio at time step t.
\item $\mathbf{h_t} = \{h_t^i | i \in \mathcal{N}\} = \{h_t^0, h_t^1,..., h_t^\mathcal{N}\} \in \mathbb{Z}^{\mathcal{N}}_+$: The number of shares owned for each asset i in $\mathcal{N}$ at time step t.
\item $\mathbf{C^i_t} \in \mathbb{R}^{\mathcal{N}}_+$: The close price of asset i in $\mathcal{N}$ at time step t.
\item $\mathbf{SS^i_t} \in (-1,0,1)$: An integer 1, 0 or -1 to indicate the sentiment of the news related to stock i at time step t.  
\item $\mathbf{T^i_t}$: The 10 different \textit{Technical Indicators} vector for asset i in the portfolio at time step t using the past prices of the asset in a specified look-back window (most common window  is 14 or 9).
\end{itemize}
To demonstrate the state space, let's assume that we have 3 different assets ($\mathcal{N} = 3$) in the trading environment and an initial capital of 1000\$ to be invested, the state vector would be a 40-dimensional vector and the \textit{initial state($s_0$)} given by the environment would be: 
$$s_0 = [[1000, 0, 0, 0][(p^1_0,  SS^1_0, T^1_0), (p^2_0,  SS^2_0, T^2_0), (p^3_0,  SS^3_0, T^3_0)]]$$

\subsection{Action Space}
The designed agent in this study receives the state $s_t$ at each time step t as input and sends back action in the range between 1 and -1 inclusive, $a_t \in [-1, 1]$, the action then is re-scaled using a constrain $K_{max}$, which represents the maximum allocation (buy/sell shares), transforming $a_t$ to an integer $K \in [-K_{max},....,-1, 0, 1, ...., K_{max}]$, which stands for the number of shares to be executed, resulting in decreasing, increasing or holding of the current position of the corresponding asset \cite{practical_DRL}. There are two important conditions regarding the action execution in our approach:
\begin{itemize}
\item If the current capital (cash) in the portfolio is insufficient to execute the buy action, the action will be partially executed with what the current capital can buy of the requested stock.
\item If the number of shares for a specific asset $(h_t^i)$ in the portfolio is less than the number of shares to be sold ($a_t^i \in \mathbb{Z}^-$), the agent will sell all the remaining shares of this asset in the portfolio.
\end{itemize}
We can mathematically express the action space as the following:
\begin{equation}
\label{action_vector}
A_t = \{a_t^i | i \in \mathcal{N}\} = \{a_t^0, a_t^1, ... , a_t^{\mathcal{N}}\}
\end{equation}
$$S.t.\,\,\,\,\,\,\,\,\,\,\,\,\,\,\,\,\,\,\,\,\,\,\,\,\,\,\,\,\,\,\,\,\,\,\,\,\,\,\,\,\,\,\,\,\,\,\,\,\,\,\,\,\,\,\,\,\,\,\,\,\,\,$$
$$a_t^i \in \mathbb{Z}^{\mathcal{N}}$$
$$-K_{max} \leq a_t^i \leq K_{max},\,\, \forall i \in \mathcal{N}$$
$$\,\,a_t^i = h_t^i\,\,\, if \,\, |a_t^i| > h_t^i\,,\, \, \forall a_t \in \mathbb{Z}^-$$
Where:
	\begin{itemize}
		\item[*] $\mathcal{N}$: assets in the portfolio.
		\item[*] $A_t$: the action vector sent by the agent to the environment.
		\item[*] $a_t^i$: the action (number of shares) to buy/sell for asset i at time step t.
		\item[*] $K_{max}$: the maximum number of shares the agent can re-allocate of an individual asset at each time step t.
		\item[*] $h_t^i$: the portfolio position (number of shares) of asset i at time step t. 
	\end{itemize}
The action space depends on the number of assets available in the portfolio $\mathcal{N}$ and it's given as $(2\times K_{max} + 1)^{\mathcal{N}}$; hence the action space increases exponentially by increasing $\mathcal{N}$.  

\subsection{Reward Function}
The difference between the portfolio value $\mathcal{V}_t$ at the end of period t and the value at the end of previous period $t-1$ represents the immediate reward $r(s, a, s^\prime)$ received by the agent after each action, and we denote the final investment return at a target time $T_f$ as $G$.
\begin{equation}
r(s, a, s^\prime) = \mathcal{V}_t - \mathcal{V}_{t-1}
\end{equation}

Where the portfolio value $\mathcal{V}$ at each time step is calculated as:
\begin{equation}
\label{portfolio_value_fun}
\mathcal{V}_t = b_{t} + h_{t}. C_{t} 
\end{equation}
Where:
	\begin{itemize}
		\item[*] $b_t$: the available cash balance in the portfolio at time step t.
		\item[*] $h_t= \{h_t^i | i \in \mathcal{N}\}$: the position vector (number of shares of each asset) at time step step t.
		\item[*] $C_t = \{C_t^i | i \in \mathcal{N}\}$: the closing price of each asset in the portfolio at time step t.
	\end{itemize}

The transition cost can be represented in many different ways in real life, and it varies from one broker to another. To better simulate the real-world trading process in the stock market, transaction costs (i.e., commission fees) are incorporated into the immediate reward ($r(s, a, s^\prime)$) calculation. In this study, we set the commission as a fixed percentage of the total closed deal cash amount, where $d_{buy}$ represents the commission percentage when buying is performed, and $d_{sell}$ is the commission percentage for selling:
$$d_t = \{d_t^i | i \in \mathcal{N} \} = [d_t^0, d_t^1, ..., d_t^\mathcal{N}]$$
\begin{equation*}
where:
d_t^i = 
\begin{cases}
    d_{buy} ,	&  \text{if } a_t^i > 0\\
    0,	& \text{if } a_t^i = 0\\
    d_{sell},	& \text{if } a_t^i < 0
\end{cases}
\end{equation*}
The commission vector $d_t$ is incorporated into the immediate reward function by excluding the commission amount paid from the portfolio value calculated in Eq. \ref{portfolio_value_fun}, so the agent would avoid excessive trading that results in a high commission rate and therefore avoids a negative reward:
\begin{equation}
\mathcal{V}_t = b_t + h_t. C_t - h_t. (C_{t-1} \circ d_t) 
\end{equation}
In the above equation, the amount paid for the commission is calculated by taking the Hadamard product of the commission vector $d_t$ and the closing price of the previous period $C_{t-1}$, that's because the action of buying/selling occurred on the previous state and therefore commission should be calculated using the closing prices on that state.

\subsection{Environment Constraints and Assumptions}
We impose the following constraints and assumptions on the MDP environment for two main reasons. First, to idealize and simplify the complex financial market systems (e.g., via liquidity assumption) without losing the nature of the problem, and the second reason is to make the model closer to a real-world situation.
\subsubsection{Non-Negative Balance Constraint}  The cash balance in any state is not allowed to be negative, $b_t > 0$. Therefore, the actions should not result in a negative cash balance, to achieve that, the environment prioritizes the execution of sell actions ($a_t < 0$) in the action vector $A_t$  (Eq. \ref{action_vector}) to guarantee cash liquidity in the portfolio so buy actions ($a_t > 0$) would be fulfilled afterward. If the buy action still results in a negative balance (i.e., not enough cash to fulfill the action), it is fulfilled partially with what remains in the portfolio's cash balance.

\subsubsection{Short-Selling Constraint} Short selling is prohibited in the designed environment, all portfolio's positions must be strictly non-negative:
$$\mathbf{h_t} = \{h_t^i | i \in \mathcal{N}\} = \{h_t^0, h_t^1,..., h_t^\mathcal{N}\} \in \mathbb{Z}^{\mathcal{N}}_+$$

\subsubsection{Zero Slippage Assumption} When the market volatility is high; slippage occurs between the price at which the trade was ordered and the price at which it’s completed \cite{slippage}. In this study, the market liquidity is assumed high enough to meet the transaction at the same price when it was ordered \cite{jiang2017deep}. This assumption is mostly valid in a real-world trading environment when trading in big stock markets.

\subsubsection{Zero Market Impact} In financial markets, a market participant impacts the market when it buys or sells an asset which causes the price change. The impact provoked by the agent in this study is assumed to have no effect on the market when it performs its actions. This assumption is mostly true even in real-life trading when the market volume is big enough to make the individual investment insignificant \cite{jiang2017deep}.

\section{DETAILS OF IMPLEMENTATION}
\subsection{The Trading Agent}
\label{TD3_sect}

Actor-Critic-based algorithms successfully solved the continuous action space utilizing function approximation and policy gradient methods. One of the most famous actor-critic, off-policy algorithms is the Deep Deterministic Policy Gradient algorithm (DDPG) \cite{DDPG_ref}. Still, despite the excellent performance DDPG achieved in continuous control problems, it has a significant drawback similar to many RL algorithms, which is the overestimation of action values ($\max_\textbf{a}Q(s_{t+1}, a_{t+1}$)) as a result of function approximation error. This overestimation bias is unavoidable in RL as we use estimates instead of ground truth in the learning process. In this study, as our problem has a continuous space of actions, we use Twin Delayed Deep Deterministic Policy Gradient (TD3) \cite{TD3} algorithm, which is a direct successor of DDPG but with improvements to tackle the overestimation problem. TD3 can reduce the overestimation bias, thus reducing the accumulation of errors in the learning process by introducing three main components to DDPG: 
\begin{enumerate}
\item \textit{Clipped Double Critic Networks}: The first component added is a novel clipped variant of Double Q-learning \cite{Double_Q} to replace the single critic. Using two different and separate critic networks to make an independent estimate of the value function can be used to make unbiased estimates of the actions selected using the opposite value estimate. TD3 uses a clipped double Q-learning instead of the traditional one used in Double Q-learning where it takes the smallest value of the two critic networks estimates, that is, if we use the traditional Double Q-learning in actor-critic methods, the policy and target networks are updated so slowly that they make similar estimates and offered slight improvement.

\item \textit{Delayed Updates}: The second component is added to solve the residual error accumulation formed due to the learning process without a fixed target (estimates instead). In Critic-Actor methods, this accumulation of errors is amplified due to the interaction between the policy (actor) and value (critic) networks, where the policy gradient is maximized over the value estimate. Delaying the policy network update, i.e., updating it less frequently than the value network, allows the value network to stabilize before it can be used to update the policy gradient. This results in a lower variance of estimates and, therefore, better policy.

\item \textit{Target Policy Smoothing Regularization:} The final component is applying a regularization strategy to the target policy by adding a small random noise and averaging over mini-batches. This is important to reduce the variance of the target values when updating the critic, which causes by overfitting spikes in the value estimate.
\end{enumerate}

\begin{algorithm}[h]
\DontPrintSemicolon
\renewcommand{\thealgocf}{}
1. \textbf{Initialization}\;
	Critic networks $Q(s,a|w_1)$, $Q(s,a|w_2)$ and actor $\pi(s|\theta)$, randomly, with weights $W_1, W_2$ and $\theta$.\;
	Target networks $Q_{1}^{\prime}$, $Q_{2}^{\prime}$ and $\pi^{\prime}$ with weights $W_{1}^{\prime} \longleftarrow W_{1}, W_{2}^{\prime} \longleftarrow W_{2},  \theta^{\prime} \longleftarrow \theta$\;
  Replay buffer $\mathcal{D}$
	
\BlankLine

2. \ForEach{t=1 to T}{
	Initialize a random process $N$ for action exploration\;
	Select action with exploration noise $a \sim \pi(s|\theta) + \epsilon, \epsilon \sim \mathcal{N}(0, \sigma)$\;
	Observe reward r and next state $s^{\prime}$\;
	Store transition tuple (s, a, r, $s^{\prime}$) in $\mathcal{D}$\;
	
	Sample mini-batch of N transitions (s, a, r, $s^{\prime}$) from $\mathcal{D}$\;
	$\tilde{a} \leftarrow \pi(s^{\prime}|\theta) + \epsilon, \epsilon \sim clip(\mathcal{N}(0, \tilde{\sigma}), -c, c)$\;
	$y \leftarrow r + \gamma \min_{i=1,2}Q(s^{\prime},\tilde{a}|w_{i})$\;
	Update critics $W_{i} \leftarrow \arg\min_{W_i}N^{-1} \sum(y-Q_{W_i}(s,a))^2$\;
	\If {t mode d}{
		Update $\theta$ by the deterministic policy gradient:\;
		$\nabla_{\theta} J(\theta) = N^{-1} \sum{\nabla_{a} Q_{W_1}(s,a)|_{a=\pi_{\theta}(s)}} \nabla_{\theta}\pi_{\theta}(s)$\;
		Update target networks:\;
		$W_{i}^{\prime} \leftarrow \tau W_{i}+(1-\tau) W_{i}^{\prime}$\;
		$\theta^{\prime} \leftarrow \tau \theta+(1-\tau) \theta^{\prime}$
	}

	}
\caption{Twin Delayed Deep Deterministic Policy Gradient (TD3) \cite{TD3}}
\label{TD3_algo}
\end{algorithm}
The agent in this paper performs daily trading operations and to aid the agent to understand its environment (the stock market), we augmented the state representation of ten different technical indicators and news sentiment scores.
\subsection{Technical Indicator}
We used the ten most famous indicators used by technical traders when trading in the stock market \cite{technical_analysis}, we describe them briefly as follows:
\begin{enumerate}
		 \item \textbf{Relative Strength Index (RSI)} $\in \mathbb{R}^{\mathcal{N}}_+$: A momentum indicator to measure the magnitude of recent price changes and identify overbought or oversold conditions in the stock price.
		 
		 \item \textbf{Simple Moving Average (SMA)}  $\in \mathbb{R}^{\mathcal{N}}_+$: An important indicator to identify current price trends and the potential for a change in an established trend.
		 \item \textbf{Exponential Moving Average (EMA)} $\in \mathbb{R}^{\mathcal{N}}_+$: Like SMA, EMA is a technical indicator used to spot current trends over time. However, EMA is considered an improved version of SMA by giving more weight to the recent prices considering old price history less relevant; therefore it responds more quickly to price changes than SMA.

	\item \textbf{Stochastic Oscillator (\%K)} $\in \mathbb{R}^{\mathcal{N}}_+$: It's a momentum indicator comparing the closing price of the stock to a range of its prices in a look-back window period $\mathcal{W}$.
	
	\item \textbf{Moving Average Convergence/Divergence (MACD)} $\in \mathbb{R}^{\mathcal{N}}$: Is one of the most used momentum indicators to identify the relationship between two moving averages of the stock price and it helps the agent to understand whether the bullish or bearish movement in the price is strengthening or weakening \cite{RSI}. 

	\item \textbf{Accumulation/Distribution Oscillator (A/D)} $\in \mathbb{R}^{\mathcal{N}}$: A volume-based cumulative momentum indicator that helps the agent to assess whether the stock is being accumulated (bought) or distributed (sold) by measuring the divergences between the volume flow and the stock price. 
	\item \textbf{On-Balance Volume Indicator (OBV)} $\in \mathbb{R}^{\mathcal{N}}$: Another volume-based momentum indicator that uses volume flow to predict the changes in stock price \cite{granville}:

﻿	\item \textbf{Price Rate Of Change (ROC)} $\in \mathbb{R}^{\mathcal{N}}$: A momentum-based indicator that measures the speed of stock price changes over the look-back window $\mathcal{W}$. 
	
	\item \textbf{William's \%R} $\in \mathbb{R}^{\mathcal{N}}_+$: Known also as Williams Percent Range,  is a momentum indicator used to spot entry and exit points in the market by comparing the closing price of the stock to the high-low range of prices in the look-back window ($\mathcal{W}$).
	
	\item \textbf{Disparity Index} $\in \mathbb{R}^{\mathcal{N}}_+$: Its value is a percentage that indicates the relative position of the current closing price of the stock to a selected moving average. In this study, the selected moving average is the EMA of the look-back window ($\mathcal{W}$).
\end{enumerate}

\subsection{Sentiment Scores}
The supply and demand fluctuations in the stock market are highly sensitive to the moment's news due to the impact of mass media on the investor's behavior. Hence many traders and investors consider the news reports in their stock-picking strategy. In our proposed approach, we believe that incorporating the general news sentence towards the asset being considered in the observation (state) definition will help the agent learn a better trading strategy. In Ding et al. \cite{ding-etal-2014-using} study, they showed that news headlines are more useful in forecasting than using the entire news article content. Therefore, we only consider news headlines as our input to calculate the sentiment score.
We describe the process of calculating a sentiment score for each asset in the portfolio at time step t (day) as the following:
\begin{itemize}
\item We use a rule-based matching approach to search for the asset name, stock symbol, or other keywords in the headline news (ex. Microsoft or MSFT, tech,..) released on day t.
\item Then we use a fine-tuned BERT model called FinBERT \cite{finBERT} to calculate the sentiment probability (Positive, Negative, or Neutral) of each news headline. FinBERT model is a pre-trained NLP model to analyze sentiments specifically for financial text.
\item Finally, we take the average of the asset's news sentiment probabilities for each day and assign 1 if the positive probability is higher than the negative probability and -1 otherwise. We ignore the neutral probability as we believe that if an asset has been mentioned on the news, it will impact the asset price (positively or negatively). If the asset has no news on a given day, we assign 0 to the sentiment score.
\end {itemize}

\section{EXPERIMENTS AND RESULTS}
We evaluate our proposed approach by performing two different back-testing which is the process used by traders and analysts to assess the viability of a trading strategy by testing it on historical data.

We perform two different back-testing experiments, the purpose of the first experiment (Section. \ref{first_exp}) is to validate the superiority of the continuous action space to solve the trading problem by comparing the results of the same experiments reported by Kaur \cite{kaur}. In their paper a discrete action space is adapted to solve the problem, where the agent can choose to buy, sell or hold action (i.e., discrete action space) of a fixed number of shares on each time step for a portfolio of two assets, namely; Qualcomm (QCOM) and Microsoft (MSFT). We back-test our approach on the same 5-years daily historical stock data (between 2011-2016) used in their study with the same amount of initial capital (\$10,000).

The second experiment (Section. \ref{second_exp}) is conducted to validate the robustness of our model on large space of actions and states by considering \textbf{10} different assets in the portfolio. In addition, considering that the first experiment was on training data set only, we evaluate the performance on an unseen market data (test data set) to check the agent's ability of generalization.

We use two metrics to evaluate our results: the first metric is the \textit{cumulative sum of reward}, i.e., the total profits at the end of the trading episode. The second metric is the annualized \textit{Sharpe ratio} \cite{Sharpe49} that combines the return and the risk to give the average of the risk-free return by the portfolio's deviation. In general, a Sharpe ratio above 1.0 is considered to be \textit{``good''} by investors because this suggests that the portfolio is offering excess returns relative to its volatility. A Sharpe ratio higher than 2.0 is rated as \textit{``very good''} where a ratio above 3.0 is considered \textit{``excellent''}.

\subsection{Data Description and Preprocessing}
In this work, We use Yahoo Finance \cite{yahoo} to retrieve historical market daily prices. The retrieved historical data consists of 7 columns; \textit{Date, Volume, Open, Close, Adjusted Close, High} and \textit{Low} prices. To prepare each dataset to be used by the model, we first perform timestamps processing by using the trading calendar (exchange-calendars package \cite{exchange-calendars}) to check if the market was open on the given dates to the agent and exclude weekends and holidays from the dataset so the agent will not face gaps in the trading process. Further dataset processing is required to ensure that all financial assets (stocks) considered in the portfolio have an equal length of historical data points. Some stocks have been recorded for decades, while other newly listed stocks are only a few months. This time-dimension alignment of stocks' historical data will prevent the bias action of the agent towards the stock with more data. Once we have the timestamps processed we use \textbf{Close, High, Low} prices and \textbf{Volume} at each timestamp to calculate the technical indicators of each asset with a look-back window ($\mathcal{W}$).

To obtain a comprehensive and accurate financial news, we combined headline news from \textit{Benzinga, Seeking Alpha, Zacks} and other financial news websites \cite{kaggleData1}, and crawled historical news headlines from \textit{Reddit worldNews Channel} \cite{kaggleData2}. The final dataset consists of 3,288,724 news headlines ranging between 2009-2021, which we utilized to calculate the sentiment score.

\subsection{First Experiment}
\label{first_exp}
In the first experiment, we conduct three evaluations similar to the benchmark paper \cite{kaur}. All three evaluations share the same configurations like the number of assets in the portfolio, initial capital, commission rates, etc. but with different components of the environment's state representation. We start with a baseline which only contains the close price as a market signal feature, we then add technical indicators in the second evaluation, and finally, we evaluate by adding sentiment analysis scores. In Table. \ref{firs_exp_table} we summarize the three evaluations results of the experiment.

Due to the stochasticity in the learning process, the experiment results may change at each run depending on different factors such as the actions the agent randomly starts with and uses to explore or the random weight initialization. As suggested in \cite{henderson2019deep} to ensure fairness and reliability of our results, we average multiple runs over different random seeds to have an insight into the population distribution of the algorithm performance on an environment. In this experiment's evaluations, we report and highlight results across several independent runs. While the recommended number of trials to evaluate an RL algorithm is still an open question in the field, we reported the mean and standard error across five trials (runs), which is the suggested number in many studies \cite{henderson2019deep}.
\\
\subsubsection{Evaluation on Baseline Environment}
To evaluate the continuous action approach in our model, we test it by solving the problem with only the close price of the assets as a market signal; hence the state representation in this baseline environment consists of only the \textit{position state} and the close price of the asset at t (\textbf{$C_{t}$}) as a market signal, i.e., the agent will solely make its trading decisions based on merely the closing price of the stock as a market feature. We perform 5 experiment trials each with 200 epochs (episodes) for the same hyperparameter configuration, only varying the random seed across trials.

\begin{figure*}[h]
\begin{center}
\hspace*{-2cm} 
\includegraphics[width=22cm,height=8cm]{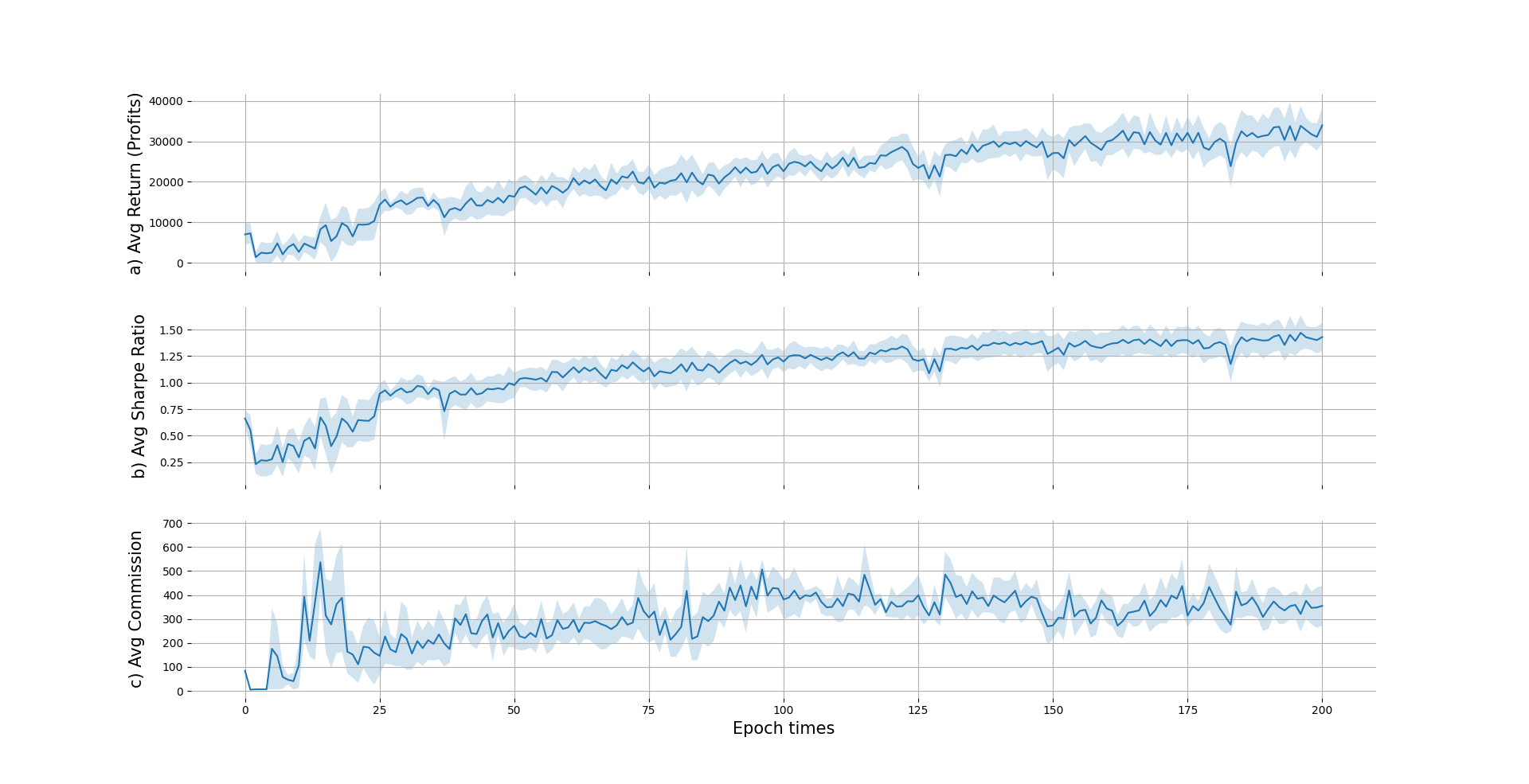}
\end{center}
\vspace*{-0.9cm}
\caption{TD3 agent performance metrics on Baseline environment using the same hyperparameter
configurations averaged over 5 different random seeds. \textbf{a)} Average return (Profits in dollars) at the end of each episode. \textbf{b)} The average annual Sharpe ratio at the end of each episode. \textbf{c)} The average amount of commission spent at the end of each episode}
\label{baseline_figs}
\end{figure*}

Fig. \ref{baseline_figs} shows the average return (sum of rewards) at each trading episode and the standard error across the 5 runs. As can be observed, the agent's performance increases with more experience it gains with the number of epochs to successfully achieve 33960\$ average return (profits) with standard error equals to $\pm{4473}$\$. From the commission spent by the agent, we can conclude that the agent was successfully able to find a balanced trading strategy by balancing between trading and holding positions. Finally, the average annual Sharpe ratio of our approach on the baseline environment was 1.43 with a standard error of $\pm{0.13}$. This is significantly higher than the reported Sharpe ratio 0.85 in \cite{kaur} benchmark.
\\
\subsubsection{Evaluation on WithTechIndicators Environment}
Using the same configurations used in baseline environment evaluation, we augment the state with technical indicators and run 5 independent experiments to report the average return, Sharpe ratio, and commission.
We refer to this environment with technical indicators and close price in the state representation as \textit{WithTechIndicators} environment.

\begin{figure}[htp]
\begin{center}
\includegraphics[width=\linewidth]{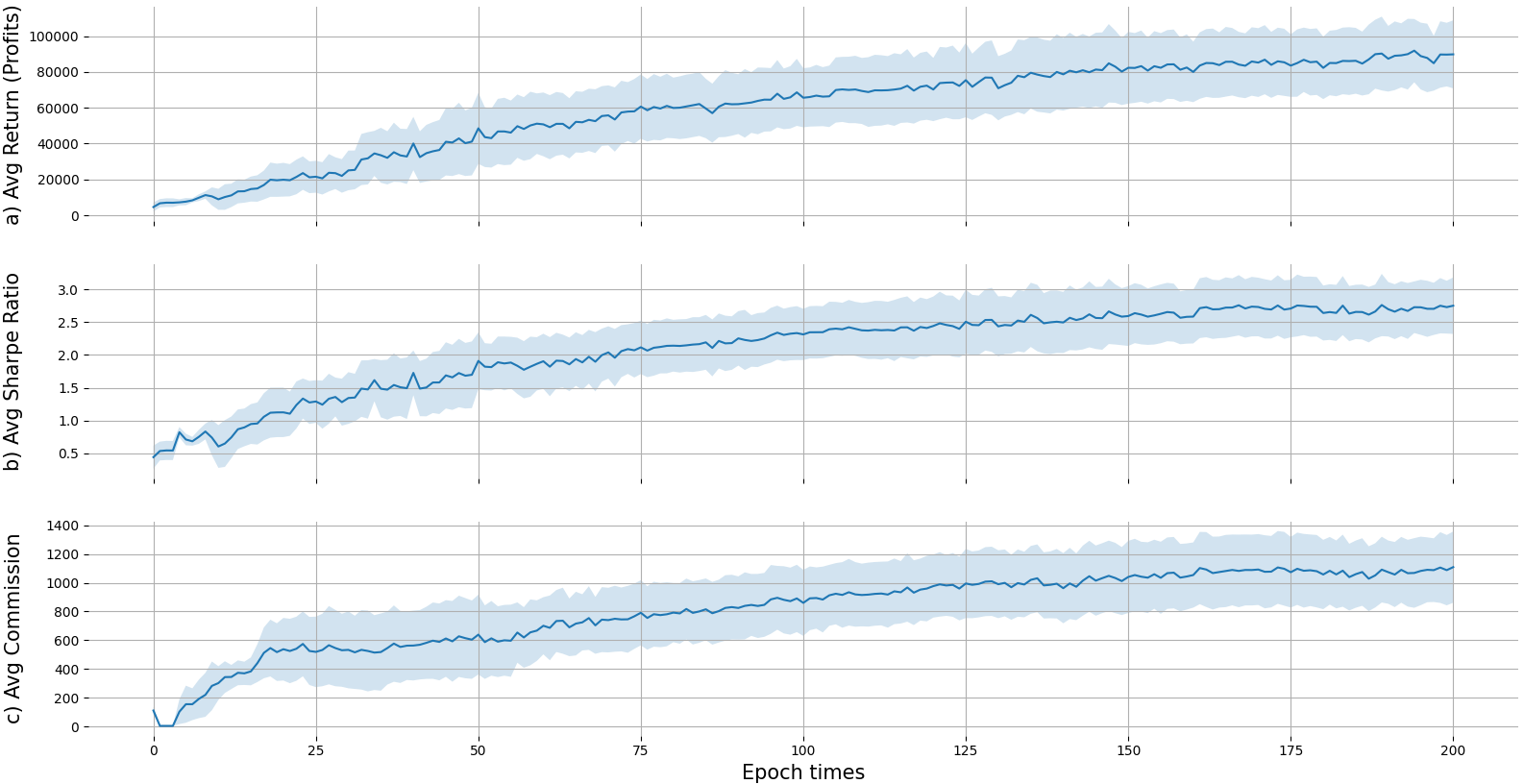}
\end{center}
\vspace*{-0.2cm}
\caption{TD3 agent performance metrics on WithTechIndicators Environment using the same hyperparameter
configurations averaged over 5 different random seeds. \textbf{a)} Average return (Profits in dollars) at the end of each episode. \textbf{b)} The average annual Sharpe ratio at the end of each episode. \textbf{c)} The average amount of commission spent at the end of each episode}
\label{techindicators_figs}
\end{figure}

The results in Fig. \ref{techindicators_figs} demonstrate that augmenting the environment with technical indicators has brought more helpful information to the agent to make better decisions. The agent successfully achieved 89782\$ average return (profits) with $\pm{18980}$\$ standard error, and an average Sharpe ratio equals 2.75 with a standard error $\pm{0.43}$. We can also notice that the average amount of commission is almost two times the amount spent in the baseline environment, which means that the agent was significantly more active in buying/selling stocks and closed more successful deals. In addition, our approach outperformed the benchmark reported Sharpe ratio of 1.4.
\\

\subsubsection{Evaluation on WithSentiments Environment}
\begin{figure}[htp]
\begin{center}
\includegraphics[width=\linewidth]{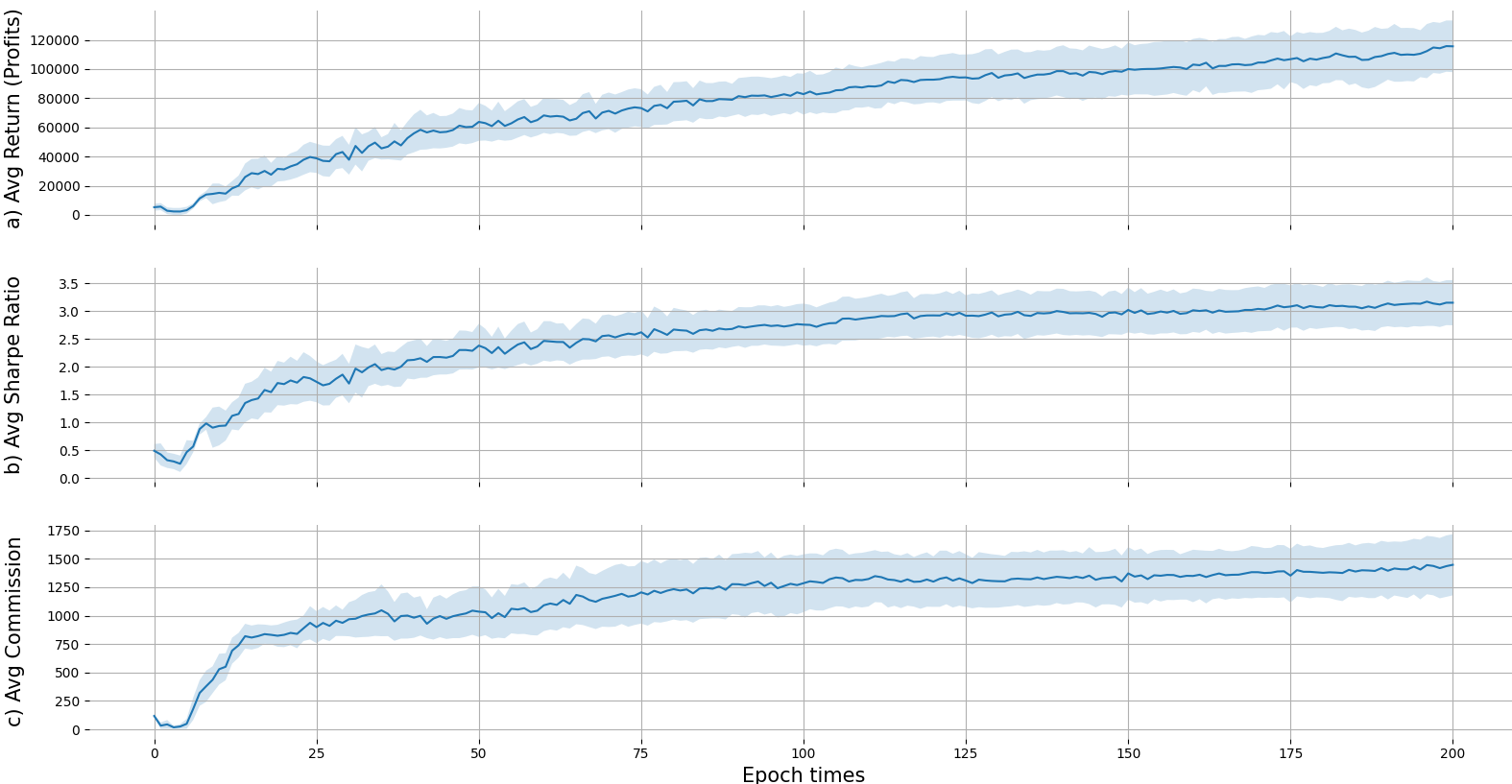}
\end{center}
\vspace*{-0.2cm}
\caption{TD3 agent performance metrics on WithSentiments Environment using the same hyperparameter
configurations averaged over 5 different random seeds. \textbf{a)} Average return (Profits in dollars) at the end of each episode. \textbf{b)} The average annual Sharpe ratio at the end of each episode. \textbf{c)} The average amount of commission spent at the end of each episode}
\label{withsentiments_figs}
\end{figure}

We refer to this environment with sentiment analysis scores, technical indicators, and close price in the state representation as \textit{WithSentiments} environment. We include the sentiment scores of news headlines for each asset in the state representation and repeat the experiment with the same configurations. The total average return profits increased to 115591\$ with standard error equals to $\pm{17721}$ across the five runs. Sharpe ratio increased to 3.14 and $\pm{0.40}$ standard error. The average amount of commission equals the amount spent in the environment with only technical indicators (WithTechIndicators environment), which means that the agent performed almost the same number of trades but with a better decision (policy). In the benchmark \cite{kaur} study, they also reported an increase in the agent performance when adding sentiment scores to the state with a Sharpe ratio equal to 2.4.
The plot showing the results in Fig. \ref{withsentiments_figs} demonstrates that augmenting the state with sentiment analysis along with technical indicators has improved the agent performance.

\textbf{Experiment's Summary}\\
We notice in all plots of the three evaluations that the policy improves over time, as the agent accumulates more reward, and thus the Sharp ratio increases. Towards the end the slope is almost flat indicating that the policy has stabilized to local optimum. As the stock trading problem has never been solved we do not have a specified reward or Sharpe ratio threshold at which it's considered solved.
\begin{table}[htb]
\caption{THE PERFORMANCE EVALUATION COMPARISON BETWEEN THREE DIFFERENT EVALUATIONS AND BENCHMARK}
\resizebox{\columnwidth}{1.2cm}{%
\begin{tabular}{cccc}
\hline
\textbf{Evaluation Environment} &
  \textbf{Baseline} &
  \textbf{WithTechIndicators} &
  \textbf{WithSentiments} \\ \hline
\textbf{Accumulated Return}       & 33960\$ $\pm{4473}$ & 89782\$ $\pm{18980}$ & 115591\$ $\pm{17721}$ \\ \hline
\textbf{Sharpe Ratio} & 1.43 $\pm{0.13}$    & 2.75 $\pm{0.43}$     & 3.14 $\pm{0.4}$       \\ \hline
\textbf{Commission}       & 355\$ $\pm{83}$     & 1109\$ $\pm{248}$    & 1447\$ $\pm{268}$     \\ \hline
\textbf{\begin{tabular}[c]{@{}c@{}}Sharpe Ration\\ benchmark
\end{tabular}} &
  0.85 &
  1.4 &
  2.4 \\ \hline
\end{tabular}
}

\label{firs_exp_table}
\end{table}

\subsection{Second Experiment}
\label{second_exp}
In the second experiment, we evaluate our approach on a wider action and state spaces by considering 10 assets to trade, AAPL, MSFT, QCOM, IBM, RTX, PG, GS, NKE, DIS and AXP. 

\begin{figure}[htp]
\begin{center}
\includegraphics[width=\linewidth]{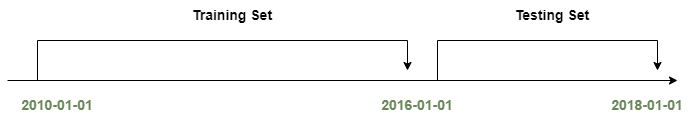}
\end{center}
\vspace*{-0.2cm}
\caption{Train, and test data splits}
\label{data_split}
\end{figure}

Our back-testing uses historical daily data from 01/01/2010 to 01/01/2018 with initial capital of 100000\$ for performance evaluation. We split the data set into two periods, the first period is to train the agent, the second is used to test the performance of the agent on unseen data  (Fig. \ref{data_split}).

We notice that for our model to generalize better, we had to impose regularization by normalizing the observation space using \textit{Batch Normalization}. This technique uses mini-batches from samples to have unit mean and variance. It maintains a running moving average of the mean and variance to normalize the observation vector during testing. We further normalized the rewards received by the agent as it makes the gradient steeper for better rewards. We also set the look-back window to 20 ($\mathcal{W} =20$).
We added action noise to encourage exploration during training to force the agent to try different actions and explore its environment more effectively, leading to higher rewards and more elegant behaviors.

Our approach successfully archived a 2.68 Sharpe ratio which considered \textit{``very good''} and 110308\$ as total profits (Rewards) on the test data. We let the agent keep learning on the test set since this will help the agent better adapt to the market dynamics.

\textbf{Disabling Sell Action:} To investigate whether profits made on the test data (between 2016 and 2018) are a matter of the standard increase in the stocks and the market growth in general (primarily that tech stocks are known for their excellent performance in the past years) or are made due to the decision made by the agent, we disable the sell action and only let the agent buy and hold during the trading episode. As a result the agent allocated the capital as follows:
\begin{table}[H]
\centering
\begin{tabular}{cc}
\hline
\textbf{Stock} & \textbf{Shares} \\ \hline
\textbf{AAPL}  & 751             \\ \hline
\textbf{AXP}   & 398             \\ \hline
\textbf{MSFT}  & 398             \\ \hline
\textbf{IBM}   & 200             \\ \hline
\textbf{DIS}   & 199            
\end{tabular}
\end{table}

and hold on to this position until the end of the episode. The Sharpe ratio has decreased to 2.00 with 66949\$ as total profits (Rewards), which indicates that the decision made by the agent had a positive effect on the return and it was not merely due to the natural growth of the market.

\section{CONCLUSION AND FUTURE WORKS}
This work presented a Deep Reinforcement Learning approach that combines technical indicators with sentiment analysis to find an optimal trading policy for assets in the stock market. Results show that the addition of technical indicators and sentiment scores of the news headlines to the state representation has significantly improved the agent's performance and the superiority of using a continuous action space over a discrete one to solve the trading problem. We also explored the potential of using an Actor-Critic algorithm (TD3) to solve the portfolio allocation problem.  Our approach achieved an annual Sharpe ratio of 2.68 on test data, which is considered "Good" by investors.
The approach can be improved in future work by having more computational power to run more experiences and better evaluate the approach. Our environment, agent, and learning process possess many hyperparameters that must be tuned. It will be interesting to see the model's performance with better-tuned parameters, which requires high computation power. In addition, we believe that training an NLP algorithm to process the financial news content instead of only the headline may positively affect the agent performance.

\ifCLASSOPTIONcaptionsoff
  \newpage
\fi



%\bibliographystyle{IEEEtran}

\begin{thebibliography}{1}

%%%% SL
\bibitem{PATEL2015259}
J.~Patel, S.~Shah, P.~Thakkar, and K.~Kotecha, ``Predicting stock and stock
  price index movement using trend deterministic data preparation and machine
  learning techniques,'' {\em Expert Systems with Applications}, vol.~42,
  no.~1, pp.~259--268, 2015.

\bibitem{8010701}
A.~Tsantekidis, N.~Passalis, A.~Tefas, J.~Kanniainen, M.~Gabbouj, and
  A.~Iosifidis, ``Forecasting stock prices from the limit order book using
  convolutional neural networks,'' in {\em 2017 IEEE 19th Conference on
  Business Informatics (CBI)}, vol.~01, pp.~7--12, 2017.
  
\bibitem{10.1371/journal.pone.0234107}
A.~Ntakaris, J.~Kanniainen, M.~Gabbouj, and A.~Iosifidis, ``Mid-price
  prediction based on machine learning methods with technical and quantitative
  indicators,'' {\em PLOS ONE}, vol.~15, pp.~1--39, 06 2020.

\bibitem{app10113961}
Y.~Hao and Q.~Gao, ``Predicting the trend of stock market index using the
  hybrid neural network based on multiple time scale feature learning,'' {\em
  Applied Sciences}, vol.~10, no.~11, 2020.
  
%%%%

\bibitem{LopezdePrado2018The1R}
M.~M.~L. de~Prado, ``The 10 reasons most machine learning funds fail,'' {\em
  WGSRN: Data Collection \& Empirical Methods (Topic)}, 2018.
  
\bibitem{Meng}
T.~L. Meng and M.~Khushi, ``Reinforcement learning in financial markets,'' {\em
  Data}, vol.~4, no.~3, 2019.

\bibitem{MDP_book}
M.~L. Puterman, {\em Markov decision processes: discrete stochastic dynamic
  programming}.
\newblock John Wiley \& Sons, 2010.

\bibitem{chakraborty2019capturing}
S.~Chakraborty, ``Capturing financial markets to apply deep reinforcement
  learning,'' 2019.

%%% RL with desecrate action space
\bibitem{8489208}
M.~R. Vargas, C.~E.~M. dos Anjos, G.~L.~G. Bichara, and A.~G. Evsukoff, ``Deep
  leaming for stock market prediction using technical indicators and financial
  news articles,'' in {\em 2018 International Joint Conference on Neural
  Networks (IJCNN)}, pp.~1--8, 2018.

\bibitem{Bertoluzzo2014}
M.~Corazza and F.~Bertoluzzo, ``Q-learning-based financial trading systems with
  applications,'' Working Papers 2014:15, Department of Economics, University
  of Venice "Ca' Foscari", 2014.

\bibitem{TAN20114741}
Z.~Tan, C.~Quek, and P.~Y. Cheng, ``Stock trading with cycles: A financial
  application of anfis and reinforcement learning,'' {\em Expert Systems with
  Applications}, vol.~38, no.~5, pp.~4741--4755, 2011.

\bibitem{Deng}
Y.~Deng, F.~Bao, Y.~Kong, Z.~Ren, and Q.~Dai, ``Deep direct reinforcement
  learning for financial signal representation and trading,'' {\em IEEE
  Transactions on Neural Networks and Learning Systems}, vol.~28, no.~3,
  pp.~653--664, 2017.
%%%%%%

%%% BAckground and related work
\bibitem{alagoz_hsu_schaefer_roberts_2009}
O.~Alagoz, H.~Hsu, A.~J. Schaefer, and M.~S. Roberts, ``Markov decision
  processes: A tool for sequential decision making under uncertainty,'' {\em
  Medical Decision Making}, vol.~30, no.~4, p.~474–483, 2009.
  
\bibitem{RL_book}
R.~S. Sutton, F.~Bach, and A.~G. Barto, {\em Reinforcement Learning: An
  Introduction}.
\newblock MIT Press Ltd, 2018.

\bibitem{bellman}
R.~E. Bellman, {\em Dynamic programming}.
\newblock Princeton University Press, 2010.

\bibitem{DQL_paper}
V.~Mnih, K.~Kavukcuoglu, D.~Silver, A.~Graves, I.~Antonoglou, D.~Wierstra, and
  M.~Riedmiller, ``Playing atari with deep reinforcement learning,'' 2013.
  
\bibitem{Mnih_2015}
V.~Mnih, K.~Kavukcuoglu, D.~Silver, A.~Rusu, J.~Veness, M.~Bellemare,
  A.~Graves, M.~Riedmiller, A.~Fidjeland, G.~Ostrovski, S.~Petersen,
  C.~Beattie, A.~Sadik, I.~Antonoglou, H.~King, D.~Kumaran, D.~Wierstra,
  S.~Legg, and D.~Hassabis, ``Human-level control through deep reinforcement
  learning,'' {\em Nature}, vol.~518, pp.~529--33, 02 2015.
  
\bibitem{Double_Q}
H.~van Hasselt, A.~Guez, and D.~Silver, ``Deep reinforcement learning with
  double q-learning,'' {\em CoRR}, vol.~abs/1509.06461, 2015.
  
\bibitem{lillicrap2019continuou}
T.~P. Lillicrap, J.~J. Hunt, A.~Pritzel, N.~Heess, T.~Erez, Y.~Tassa,
  D.~Silver, and D.~Wierstra, ``Continuous control with deep reinforcement
  learning,'' 2019.

\bibitem{TD3}
S.~Fujimoto, H.~van Hoof, and D.~Meger, ``Addressing function approximation
  error in actor-critic methods,'' {\em CoRR}, vol.~abs/1802.09477, 2018.
  
\bibitem{Bertoluzzo2012}
F.~Bertoluzzo and M.~Corazza, ``Testing different reinforcement learning
  configurations for financial trading: Introduction and applications,'' {\em
  Procedia Economics and Finance}, vol.~3, pp.~68--77, 2012.
\newblock International Conference Emerging Markets Queries in Finance and
  Business, Petru Maior University of Tîrgu-Mures, ROMANIA, October 24th -
  27th, 2012.
  
\bibitem{beating_the_stock_market}
L.~Conegundes and A.~C.~M. Pereira, ``Beating the stock market with a deep
  reinforcement learning day trading system,'' in {\em 2020 International Joint
  Conference on Neural Networks (IJCNN)}, pp.~1--8, 2020.
%%%%%%
%%%%%%% Problem Description
\bibitem{technical_analysis}
C.~Kirkpatrick and J.~R. Dahlquist, ``Technical analysis: The complete resource
  for financial market technicians,'' 2006.

\bibitem{fundamental_analysis}
J.~R. Nofsinger, ``The impact of public information on investors,'' {\em
  Journal of Banking \& Finance}, vol.~25, no.~7, pp.~1339--1366, 2001.
  
\bibitem{practical_DRL}
Z.~Xiong, X.-Y. Liu, S.~Zhong, H.~Yang, and A.~Walid, ``Practical deep
  reinforcement learning approach for stock trading,'' 2018.
  
\bibitem{slippage}
``{Investopedia} -- slippage definition.''
  \url{https://www.investopedia.com/terms/s/slippage.asp}.
\newblock [Online; accessed 02-October-2021].

\bibitem{jiang2017deep}
Z.~Jiang, D.~Xu, and J.~Liang, ``A deep reinforcement learning framework for
  the financial portfolio management problem,'' 2017.
%%%%%%% 

%%% The trading agent
\bibitem{DDPG_ref}
A.~Akhmetzyanov, R.~Yagfarov, S.~Gafurov, M.~Ostanin, and A.~Klimchik,
  ``Continuous control in deep reinforcement learning with direct policy
  derivation from q network,'' in {\em Human Interaction, Emerging Technologies
  and Future Applications II}, (Cham), pp.~168--174, Springer International
  Publishing, 2020.
\bibitem{RSI}
T.~T.-L. Chong, W.-K. Ng, and V.~K.-S. Liew, ``Revisiting the performance of
  macd and rsi oscillators,'' {\em Journal of Risk and Financial Management},
  vol.~7, no.~1, pp.~1--12, 2014.
  
\bibitem{granville}
J.~Granville, {\em Granville’s New Key to Stock Market Profits}.
\newblock Papamoa Press, 2018.

\bibitem{ding-etal-2014-using}
X.~Ding, Y.~Zhang, T.~Liu, and J.~Duan, ``Using structured events to predict
  stock price movement: An empirical investigation,'' in {\em Proceedings of
  the 2014 Conference on Empirical Methods in Natural Language Processing
  ({EMNLP})}, (Doha, Qatar), pp.~1415--1425, Association for Computational
  Linguistics, Oct. 2014.

\bibitem{finBERT}
D.~Araci, ``Finbert: Financial sentiment analysis with pre-trained language
  models,'' 2019.
%%%%%%%%%%%%%%
%%%%%% Experiments and results

\bibitem{yahoo}
``Yahoo finance.'' https://finance.yahoo.com/.

\bibitem{exchange-calendars}
G.~Manoim, ``exchange-calendars.''
  \url{https://pypi.org/project/exchange-calendars/}.

\bibitem{kaggleData1}
``{Kaggle} -- daily financial news for 6000+ stocks.''
  \url{https://www.kaggle.com/miguelaenlle/massive-stock-news-analysis-db-for-nlpbacktests}.
\newblock [Online; accessed 15-November-2021].

\bibitem{kaggleData2}
``{Kaggle} -- sun, j. (2016, august). daily news for stock market prediction.''
  \url{ https://www.kaggle.com/aaron7sun/stocknews}.
\newblock [Online; accessed 15-November-2021].

\bibitem{henderson2019deep}
P.~Henderson, R.~Islam, P.~Bachman, J.~Pineau, D.~Precup, and D.~Meger, ``Deep
  reinforcement learning that matters,'' 2019.

\bibitem{Sharpe49}
W.~F. Sharpe, ``The sharpe ratio,'' {\em The Journal of Portfolio Management},
  vol.~21, no.~1, pp.~49--58, 1994.
  
\bibitem{kaur}
S.~Kau, ``Algorithmic trading using reinforcement learning augmented with
  hidden markov model. working paper, stanford university.,'' 2017.
%%%%%%%%%%%%%


\end{thebibliography}


%\vfill


\end{document}